\documentclass{arxiv_paper}

\pdfoutput=1

\usepackage{amsmath}
\usepackage{fancybox}
\usepackage{graphicx}
\usepackage{url}

\begin{document}

%
%
%


\large

\begin{titlepage}

   \date{3 May 2019}

   \title{Study of muon pairs in hadronic events with bottom tag using L3 data
          recorded near the Z resonance}

  \begin{Authlist}
Thomas Hebbeker\Instfoot{iiia}{RWTH Aachen University, 
        Physics Institute III A}
Stefan Roth\Instfoot{iiib}{RWTH Aachen University, 
        Physics Institute III B}
  \end{Authlist}

  \begin{abstract}
    A recent publication based on ALEPH $\mathrm{e}^+\mathrm{e}^-$
    data taken near the Z resonance reports
    an excess,  
    near 30 GeV,  in the di-muon mass spectrum
    for events
    containing b quarks.
    In this paper we measure the di-muon mass spectrum for L3 data with a bottom tag, recorded in the years
    1994 and 1995 at the LEP collider at CERN, 
    and compare it to Monte Carlo predictions.

  \end{abstract} 
  
\end{titlepage}

\setcounter{page}{2}

\section{Introduction}

Recently,
ALEPH $\mathrm{e}^+\mathrm{e}^-$
LEP collision data were reanalyzed \cite{heister}. 
In the event sample recorded at, or close, to
the Z resonance in the years between 1992 and 1995 the
opposite sign dimuon mass spectrum was investigated in
hadronic final states with a bottom tag. 
An excess of events was reported at $m = 30.40 \pm 0.46 \, \mathrm{GeV}$, 
with an intrinsic peak width of $\Gamma = 1.78 \pm 1.14 \; \mathrm{GeV}$,
relative to
an assumed smooth background, with
a significance of 3 $\sigma$ after correcting for the look-elsewhere-effect.

Using archived L3 events, we investigate  if  a similar  excess can be seen in the 
L3 data set recorded around the Z resonance.
Since the detectors ALEPH \cite{alephdetector} and L3 \cite{l3detector} are different
(muon angular acceptance, muon efficiency at low momenta, bottom tagging etc), we can not
make a direct comparison
between the dimuon mass spectra measured with the ALEPH 
and L3 analyses.
Nevertheless,
we implement as far as possible
similar selection cuts as used for the ALEPH data \cite{heister}
for the
L3 event selection.
We have fixed all selection criteria
and the search regions
before inspecting the final dimuon mass distribution.
We have not performed any optimization or variation of cuts, 
therefore the results presented here
are free
of
look-elsewhere effects.
The L3 data are not compared to a model-dependent phenomenological background curve
as done in Ref. \cite{heister}, 
but, rather, to a Standard Model Monte Carlo prediction.

\section{Data set and Event selection}

We use
the
$\mathrm{e}^+\mathrm{e}^-$
data recorded in
1994 and 1995, at center of mass energies between
$89$ and $93 \, \mathrm{GeV}$, corresponding to an integrated
luminosity of $80/\mathrm{pb}$. 
Earlier data could not be used because the
L3 silicon vertex detector \cite{l3vertexdetector}
was only installed in 1993 and
is needed for bottom tagging.
Note that, for the 1995 data set, the L3 muon endcap detectors
\cite{l3forwardmuondetector}
were available
and are used in this analysis as well.

The events were selected by making the following requirements: 
\begin{itemize}
\item at least two reconstructed muon candidates.
\item hadronic activity: number of calorimeter clusters $\geq 15 $
  \cite{l3z2000}.
\item no huge momentum imbalance:
  $p_{miss} = |\sum \vec p \,  | \leq 18 \, \mathrm{GeV}$ \cite{heister}, where the
sum runs over all calorimeter clusters and all muons. 
\item bottom tag, $B_{tag} \geq 1.5 $ \cite{l3btag}.
\end{itemize}
$B_{tag}$  is a weighted discriminant, combining the probabilites for
tracks to originate from the primary vertex \cite{l3btag}. 
The $B_{tag}$ cut value was chosen such
that the contamination of light quark events (u,d,s,c) in the final event sample is
predicted to be as low as  $5\%$ by the Monte Carlo simulation.

The
muon candidates have to fulfill the following criteria:
\begin{itemize}
\item number of measured
  segments in the bending plane of the the central region plus
  number of segments in the endcap regions
  of the muon detector
  equal to two or more.
\item at least one
  z measuring segment (coordinate along the beam) in the muon detectors. 
\item transverse momentum with respect to the beam axis
       of  $p_T \geq 2.5 \, \mathrm{GeV} $ \cite{heister}. 
\item distance of closest approach to the
  nominal vertex position in the plane
  perpendicular
  to the beam
  less than $30 \, \mathrm{mm}$.
\item distance to the nominal vertex position along the beam axis 
  less than $50 \, \mathrm{mm}$.
\end{itemize}
The  last two cuts are chosen such that dimuon events near the Z mass
are accepted with greater than $90\%$ probability. Most muons from bottom hadron
decays
survive
these vertex cuts because the average
decay length is well below $10 \, \mathrm{mm}$ in that case.
%
If more than two muon candidates remain, we select the two
having the highest energy.
If their charges are not of opposite sign, the event is rejected. 
In total we select 227 events. 
%
We have verified from the angular distribution of the muons that the
contamination of cosmics is negligible in this data set.

\section{Standard Model background}

In the Standard Model the most important sources
of bottom-tagged di-muon events
are, by far,  processes of type
\begin{eqnarray*}
  \mathrm{e^+}  \mathrm{e^-}  \to  Z/\gamma^* \to  \mathrm{b \, \bar b}
  \;\;\; . 
\end{eqnarray*}
The muons are produced
mostly 
in semileptonic bottom decays. 
The Monte Carlo generator JETSET 7.4 \cite{jetset} was used to generate events
of type 
\begin{eqnarray*}
  \mathrm{e^+}  \mathrm{e^-}  \to  Z/\gamma^* \to  \mathrm{q \, \bar q}
    \;\;\; , 
\end{eqnarray*}
where $\mathrm{q}$ stands for all quark flavors $\mathrm{u,d,s,c,b}$ accessible at LEP energies. 
These events were subsequently passed 
through the full L3 detector simulation.
The simulated events were reconstructed in the same
way as the measured $\mathrm{e}^+\mathrm{e}^-$ events.
The total Monte Carlo event statistics are 3.5 times higher than for the L3 data sample.

\section{Results}

In this section we
compare the number of measured and expected events in two search regions
in the distribution of the dimuon mass $m_{\mu\mu}$:
\begin{itemize}
\item peak region: $   29 \leq m_{\mu\mu} \leq  32 \, \mathrm{GeV}$
  \\ This mass window
is 
approximately centered on the ALEPH peak position
($m = 30.40 \pm 0.46 \, \mathrm{GeV}$)
and the width is about  the
width 
of the ALEPH peak ($\Gamma = 1.78 \pm 1.14 \; \mathrm{GeV}$)~\cite{heister}.
\item signal region: $   27 \leq m_{\mu\mu} \leq  34 \, \mathrm{GeV}$
  \\ This
  interval is also 
centered on the ALEPH peak position, with a width that is about twice the 
width of the ALEPH peak.
\end{itemize}
The dimuon invariant mass resolution was studied with the L3 Monte Carlo event sample.
Near
$30 \, \mathrm{GeV}$ it is approximately $1 \, \mathrm{GeV} $
( = $3 \, \%$ relative). 

\begin{figure}[htbp]
\begin{center}
\includegraphics[width=16cm,angle=0]{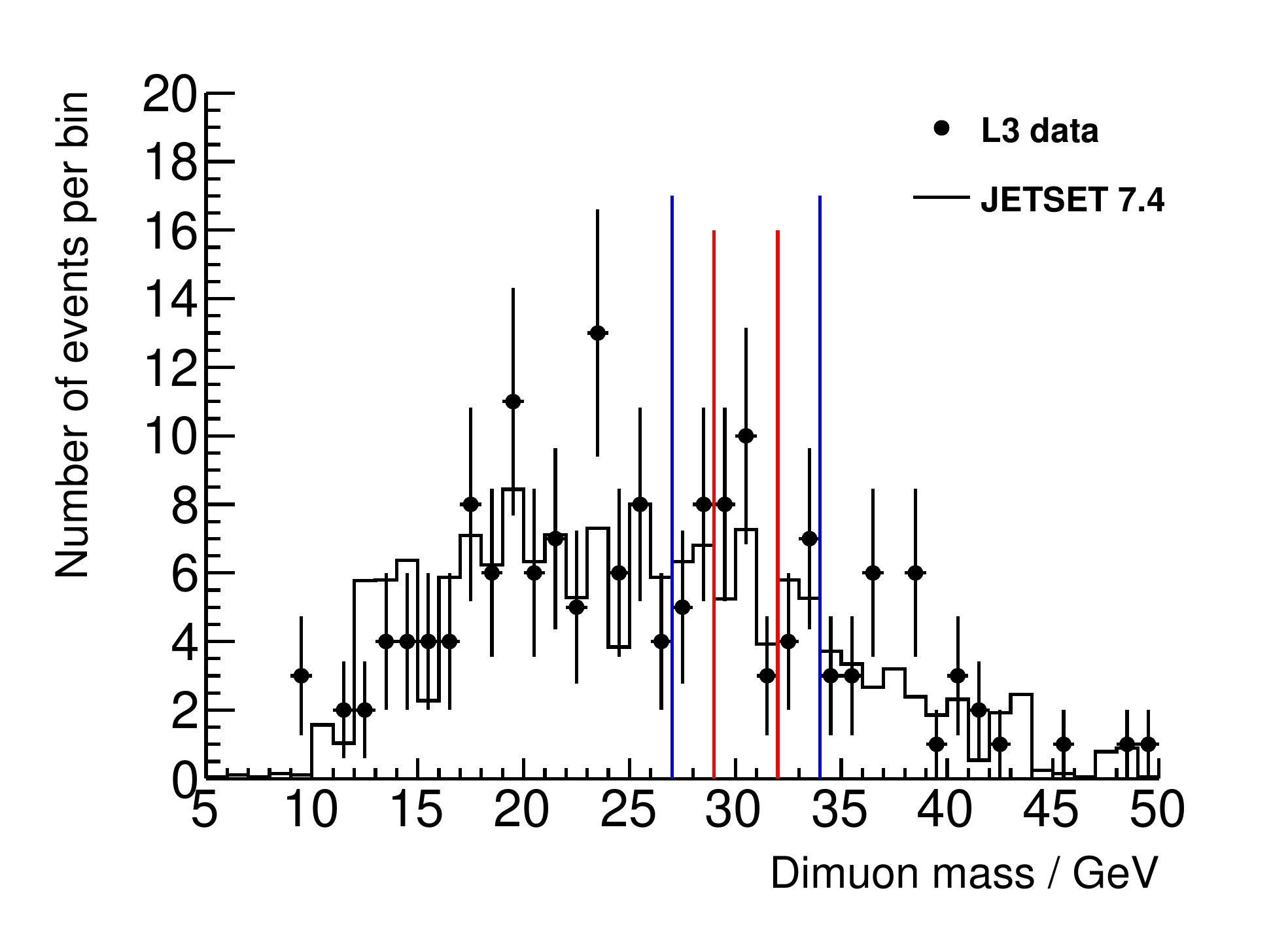} 
\caption{Opposite sign dimuon mass spectrum after all cuts, L3 data and normalized 
  Monte Carlo
  prediciton. The peak region and the signal region are marked by
  vertical red and blue lines, respectively.}
\label{mmbtag_ah}
\end{center}
\end{figure}

\begin{figure}[htbp]
\begin{center}
\includegraphics[width=16cm,angle=0]{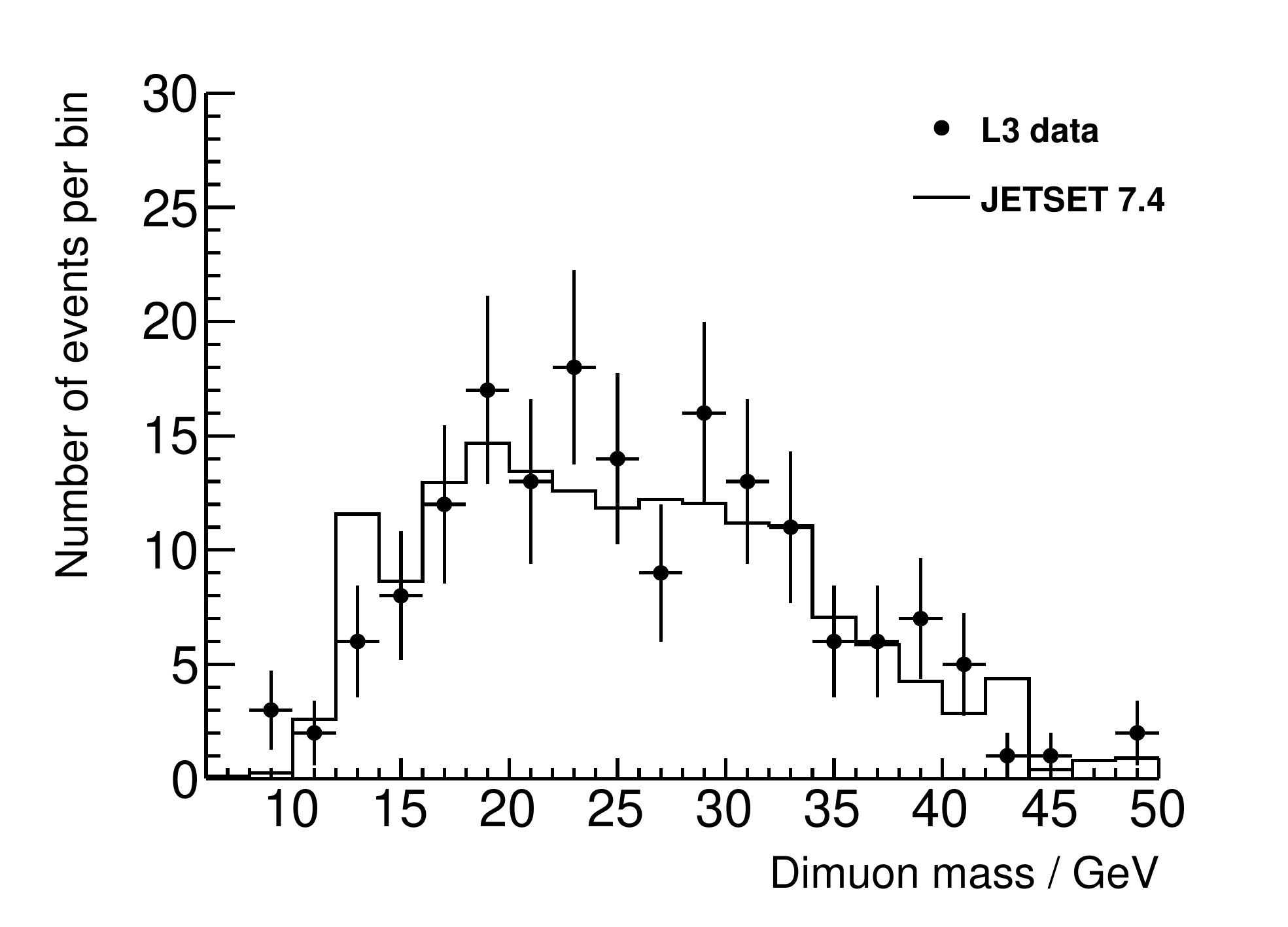} 
\caption{Dimuon mass spectrum after all cuts, L3 data and normalized
  Monte Carlo
  prediciton. The only difference with respect to
  Fig.~1 is the wider binning, $2 \, \mathrm{GeV}$
  instead of $1 \, \mathrm{GeV}$ bin width.
}
\label{mmbtag_th}
\end{center}
\end{figure}

Figure~\ref{mmbtag_ah}
shows the opposite sign dimuon invariant mass spectra
obtained from L3 data and from the JETSET Monte Carlo simulation, in the mass window
$5 - 50 \, \mathrm{GeV}$ (170 events). 
We use exactly the same binning as employed for the ALEPH analysis \cite{heister} (Figure 7).
Figure~\ref{mmbtag_th} displays the same data, though with a doubled bin width
in order to reduce the statistical fluctuations. 

The Monte Carlo histogram is normalized to the dimuon mass distribution measured from L3 data.
The signal region has been excluded in the normalization calculation.
The Monte Carlo needs to be scaled up by a factor of 1.2, relative to the normalization
according to luminosity. 
Overall the
measured opposite charge dimuon mass distribution is well reproduced by the Standard Model
Monte Carlo. 


In Figure~\ref{mmbtag_ah} both the signal region and the peak region are marked with
blue and red vertical lines.
Integrating over the corresponding bins and computing the difference:
L3 data - Monte Carlo prediction, 
we obtain the following results:
\begin{center}
  \begin{tabular}{l| c}
    region &  \#  data events -  \# MC  events  \\
    \hline
    signal ($27-34 \, \mathrm{GeV}$)   &  $ 8.4 \pm  8.6 $  \\ 
    peak $\;\,$  ($29-32 \, \mathrm{GeV}$)   & $ 6.4 \pm  7.4$ 
\end{tabular}  
\end{center}
Only statistical uncertainties are given.

\section{Summary}

We have analyzed archived L3
$\mathrm{e}^+\mathrm{e}^-$
data recorded near the Z peak. 
In the oppositely charged dimuon spectrum, measured for hadronic events with a bottom tag,
we investigated the region around the peak observed near $30 \, \mathrm{GeV}$
in ALEPH data \cite{heister}. 
We find a slight excess in the data compared to the Monte Carlo prediction, but
the deviation of the measured event yield from the Standard Model
prediction
is not statistically significant 
(about one 
standard deviation).

\section{Acknowledgements}

We thank Arno Heister for interesting discussions and Shawn Zaleski
for useful comments on the paper draft.


\bibliographystyle{unsrt}  

\end{document}